%Paper: hep-th/9205029
%From: MIRJAM@penndrls.upenn.edu
%Date: Tue, 12 May 92 21:34:18 EST
%Date (revised): Mon, 18 May 92 09:26:14 EST

%%%XXX\input iasphys

\input phyzzx

\PHYSREV
\date={May 1992}
\Pubnum={\caps UPR--505--T
%,preliminary version
 }

\titlepage
\title
{Neutrino Masses within
the Minimal Supersymmetric Standard Model}
\author{Mirjam Cveti\v c   and Paul Langacker}
\address{Department of Physics\break
University of Pennsylvania\break
Philadelphia, PA 19104--6396\break}

\abstract{
We investigate the possibility of accommodating neutrino masses
compatible with the MSW study of the Solar neutrino deficit within
the minimal supersymmetric Standard Model.  The  ``gravity-induced''
seesaw mechanism
based  on an interplay of nonrenormalizable and renormalizable
terms in the superpotential allows neutrino masses
$m_\nu\propto
m_u^2/M_I$, with $m_u$ the corresponding
quark mass and
$M_I\simeq 4\times10^{11}$ GeV, while at the same
time ensuring the grand desert with the
gauge coupling unification
at $M_U\simeq
2\times10^{16}$ GeV.  The proposed scenario may be
 realized in a class of string vacua, {\it i.e.,}  large
radius ($R^2/\alpha '={\cal O}(20)$)
$(0,2)$ Calabi-Yau spaces.
%obtained by deforming a  $(2,2)$
%Calabi-Yau space along a flat direction.
 In this case
%radius and $\alpha'$ is the string tension
 $M_U^2=M_C^2/{\cal O}
(2R^2/\alpha')$ and $M_I=
{\cal O}(e^{-R^2/\alpha'})M_C$. Here
 $M_C=g\times 5.2\times
 10^{17}$GeV
%$= {\cal O}(10^{17})$GeV
 is the
  scale of the tree level (genus zero)
 gauge coupling ($g$) unification.
}

%\pacs{To be submitted to {\sl Physical Review Letters}}
\line{PACS \# 12.10,11.15,11.17\hfill}
\endpage
\unnumberedchapters
\REF\gauge{P.~Langacker and M.~Luo, {\sl Phys. Rev.} {\bf D44}, 817
(1991); U.~Amaldi, W.~de~Boer, and H.~Furstenau, {\sl Phys. Lett.}
{\bf260B}, 447 (1991); J.~Ellis, S.~Kelley, and D.~V.~Nanopoulos,
{\sl Phys. Lett.} {\bf249B}, 441 (1990).}
\REF\MSW{S.~P.~Mikheyev and A.~Yu.~Smirnov, {\sl Yad. Fiz.} {\bf42},
1441 (1985) ({\sl Sov. J. Nucl. Phys.} {\bf42}, 913 (1985)); {\sl Nuo.
Cim.} {\bf9C}, 17 (1986); L.~Wolfenstein, {\sl Phys. Rev.} {\bf D17},
2369 (1968); {\bf D20}, 2634 (1979).}
\REF\BKL{For a recent discussion, see
S.~Bludman, D.~Kennedy, and
P.~Langacker, {\sl Phys. Rev.} {\bf D45,} 1810 (1992), and {\sl Nucl.
Phys.,} in press.}
Precise
data from the LEP experiments
indicate that the gauge couplings of the Standard Model meet at
$M_U\simeq (1-4)\times
10^{16}$ GeV in the minimal supersymmetric extension of the
Standard Model.\refmark\gauge\
  Another set of intriguing data arise from the Solar
neutrino experiments.  The deficit of Solar neutrinos can most
efficiently be explained through the MSW\refmark\MSW\
mechanism of matter-enhanced neutrino
oscillations.  In particular, current data favor\refmark\BKL\
the mass splitting of
the electron and muon neutrinos to be
$\Delta m^2 \equiv
|m_{\nu_\mu}^2-m_{\nu_e}^2|\simeq (2-5) \times 10^{-7}  {\hbox{eV}}^2$
if the mixing angle
$\theta_{\nu_\mu \nu_e}\simeq{\cal O}(\theta_C)$, where
$\theta_C$ is the Cabibbo angle. For arbitrary mixing angles the
nonadiabatic MSW solution favors
$\Delta m^2
\simeq (1-16) \times 10^{-7}  {\hbox{eV}}^2$.

In many grand unified theory (GUT)
models
 the  lepton mixing  matrix $V_{\ell}$
and the  Cabibbo-Kobayashi-Maskawa matrix $V_{CKM}$ are predicted to be
approximately
equal\Ref\footZ{
This occurs in simple GUT seesaw models for a wide range of mass
matrices for the heavy Majorana neutrinos\refmark\BKL~.}.
However,
these same models predict $m_e/m_\mu\simeq m_d/m_s$ ($\simeq1/20$), which
fails by an order of magnitude.  Small perturbations on the models which
rescue this mass relation\Ref\revw{For a review, see P.~Langacker {\sl
Phys. Rep.} {\bf72C}, 185 (1981).} may also modify the mixing
angle predictions.
In theories
with no explicit intra-family unification
$V_{CKM}$ and $V_{\ell}$ are not expected to be equal, but could
well be of the same order of magnitude.
We will assume $\sin^2 2 \theta_{\nu_\mu \nu_e} \sim \sin^2 2 \theta_C
\sim 0.18$ for the central value in our discussion, but will consider the
entire range $4 \times 10^{-3} - 1$ allowed by the nonadiabatic MSW
solution. Assuming $m_{\nu_\mu} \gg m_{\nu_e}$, the corresponding values
of  $m_{\nu_\mu}$ are $
(5-7) \times 10^{-4}  {\hbox{eV}}$ for $\theta_{\nu_\mu \nu_e}
\sim \theta_C$ and $
(3-40) \times 10^{-4}  {\hbox{eV}}$ for general $\theta_{\nu_\mu \nu_e}.$

 In the GUT
seesaw scenario\Ref\GellM{M.~Gell-Mann,
P.~Ramond, and R.~Slansky, in {\it Supergravity,} ed.
F.~van Nieuwenhuizen and
D.~Freedman (North Holland, Amsterdam 1979) p.~315; T.~Yanagida, {\sl
Prog. Th. Phys.} {\bf B135}, 66 (1978).}
masses of light neutrinos are given by
$m_{\nu_{e,\mu}}\simeq c\ m^2_{u, c}/M_I$, where $m_{u,c}
$ are the corresponding  quark masses
and $c\simeq 0.05-0.09$ is
a factor due to the renormalization down to the low energy
scale\refmark\BKL~.
This implies that
$M_I\simeq
(4 \pm 3) \times10^{11}$ GeV, the central value corresponding to
$\theta_{\nu_\mu \nu_e}
\sim \theta_C$ .
If the same scale applies to the third family,
 then $m_{\nu_\tau}\simeq
c'm_t^2/M_I$ could be in the cosmologically interesting 10 eV range,
with $\nu_\mu$--$\nu_\tau$ oscillations observable in the laboratory.

  Each of the two sets of experimental
data has an elegant theoretical explanation.  Unfortunately, the two
theoretical models are mutually exclusive at first glance.  In the
minimal supersymmetric
Standard Model, there is a ``grand desert''
up to $M_U\simeq (1-4) \times
10^{16}$ GeV.  Within this theory, the
implementation of a   naive seesaw mechanism would indicate that
$m_{\nu_{e,\mu}}
\simeq c m^2_{u,c}/M_I$, with $M_I\sim (10^{-2}-1) M_U
\sim (1-4)\times 10^{15 \pm 1}$ GeV,
which is too small to be compatible with the favored experimental
data and the MSW \break
scenario\refmark\BKL~.
Actually, in GUT models, in order to obtain a nonzero $M_I$
with $M_I\sim M_U$ one has to introduce
large Higgs representations ({\it e.g.,} the {\bf126} of $SO(10)$).

The aim of this note is to implement the neutrino masses in the minimal
supersymmetric  Standard Model in such a way that there
is still a grand desert with the
gauge coupling unification
at $M_U\sim (1-4) \times 10^{16}$GeV, while the
effective scale $M_I$ governing the neutrino masses
is in the range of  $(4 \pm 3) \times 10^{11}$ GeV.
We are proposing  a ``gravity-induced''
seesaw mechanism (an extension of a mechanism proposed by
\REF\Nandi{S. Nandi and U. Sarkar, {\it Phys. Rev. Lett}. {\bf 56},
564 (1986).}
Nandi and \break
Sarkar\refmark\Nandi~),
realized through
an interplay between the nonrenormalizable and renormalizable
terms in the superpotential,
as the origin of the neutrino  masses.
The essence of the idea is based on a supersymmetric
theory with an
extended gauge symmetry, which
contains
an additional sterile
neutrino (a Standard Model singlet), and a {\it
restricted}  representation of the Higgs fields. Such
fields
break the extended gauge symmetry
at the scale $M_U$. However, they cannot
give the sterile neutrino
a large Majorana mass  proportional to $M_U$ through
  the renormalizable (cubic) terms  in the
superpotential.  On the other
hand, through
the nonrenormalizable (\eg  , quartic)  terms
of the superpotential, which
are  suppressed by a scale  $M_{NR}>M_U$,
such Higgs fields can give a
 Majorana mass of order $M_U^2/M_{NR}<M_U$.

The origin of the nonrenormalizable terms is best motivated in
theories which include gravity, \eg,
Kaluza-Klein theories and superstring theory.
In this case the exchange of heavy modes with
masses of order the Planck mass
$M_{Pl}$ in general
induce  nonrenormalizable
terms with $M_{NR}={\cal O}(M_{Pl})$.
Thus, we shall call the proposed scenario  the
gravity-induced seesaw.
In the case of  $M_U\sim2 \times 10^{16}$ GeV and $M_{NR}\sim
M_{Pl}/\sqrt{8\pi}\sim 2 \times 10^{18}$ GeV one has
$M_I\sim 2 \times 10^{14}$ GeV,
which is about two to three
orders of magnitude too large
to be compatible with the favored MSW data.  However, if such
nonrenormalizable terms are {\it suppressed} by an additional factor
$10^{-2}-10^{-3}$, one can obtain the desired
$M_I\sim10^{11} - 10^{12}$ GeV.
%We shall
%demonstrate later how this can
%be accommodated in the study
%of superstring vacua.

Such a scenario can be accommodated within a GUT theory
with  restricted Higgs field representations; \eg, the $SO(10)$
gauge group without {\bf126}-plets of Higgs
\REF\footI{Radiative corrections due to two-loop
  effects, as proposed by E.~Witten, Phys. Lett. {\bf 91B}, 81 (1980),
  do not work in supersymmetric theory:
there are no linear terms of the type $M\phi_{10}\phi_{16}\phi_{16}$
(where {\bf10}, {\bf16}, {\bf16} are representations of the
corresponding scalar multiplets of $SO(10)$) because they break
supersymmetry.  They can be generated by soft supersymmetry
breaking terms, but this would then yield a Majorana mass of order
$M_I\sim\alpha M_W$.}
fields\refmark\footI\ . For a broad class of
simple GUT models\refmark\footZ ,
$V_{CKM} \simeq V_{\ell}$.
On the other hand,
extended gauge symmetries based on a
product of simple groups and $U(1)$ factors
(\eg, the Standard Model with additional
 $U(1)$'s  and/or
left-right symmetric gauge symmetry) with restricted
representations  of the  Higgs sector can also accommodate  the
gravity-induced seesaw mechanism. However, in this case,
the relation between $V_{CKM}$ and $V_{\ell}$
is less obvious.

One can demonstrate  the  gravity-induced seesaw
in an explicit (minimal)
model with  all the
essential features.
We  choose the enhanced
gauge symmetry,  $SU(3)_C\times
SU(2)_L\times U(1)_{Y}\times U(1)_{Y'}$, where $Y$ is the
ordinary weak hypercharge.
The matter
consists of the particle content of
the minimal Standard Model
as  well as of the
Standard Model singlets,
 $ L_i$, $S_1$ and $\bar S_1$.
$L_i$ supermultiplets with $i$ = (1,2,3)
contain  a sterile neutrino which accompanies
each of the three families.
 $S_1$ and $\bar S_1$ contain the Higgs fields
which break the enhanced gauge symmetry with
VEV's  of order $M_U$.

%The neutrino  spectrum
%for each family  consists of
%the ordinary neutrino $\nu_i
%$ and a sterile neutrino   $L_i$.
Consistent with the anomaly constraint
we choose the  following
 values for the $Y'$ charges:
quark  $SU(2)_L$ doublets,
$u_L^c$ quarks and  $e_L^c$  leptons
have (-1),
lepton   doublets  and
$d_L^c$ quarks
have (+3),  $L_i$ and $S_1$ have (-5),  $\bar S_1$ has (+5), while
Higgs doublets $H_{(1,2)}$ have (-2) and  (+2),
respectively\Ref\footII{ The choice of the particle spectrum and
the $U(1)_{Y'}$   quantum numbers are motivated  by $SO(10)$
gauge symmetry. Namely, the family particle content (including
$L_i$'s) corresponds to full {\bf 16}-plets of $SO(10)$,
the two Higgs doublets are  parts of a
 {\bf 10}-plet,
while the Standard Model singlets $S_1$ and $\bar S_1$
are  parts of a {\bf 16}-plet and a ${\bf {\overline{16}}
}$-plet, respectively.}.
In the neutrino sector,
the only renormalizable terms
allowed in the superpotential are of the
type $W=L_i\nu_i H_2$.
Terms of the type $L_i\nu_i  S_1$ or $L_iL_i S_1$
are not allowed by the quantum numbers.
These constraints yield the following  contribution to the
neutrino mass matrix:
$$
\left[\matrix{0&m\cr
             m&0\cr}\right],
\eqn\threem$$
where $m$ is proportional  to the  $VEV$ of the
Higgs doublet $H_2$. Since $H_2$ gives mass
to the quarks as well,
$m$ is of the order of the corresponding quark masses
unless there is a
large difference in the magnitude
of the  Yukawa
couplings.
On the other hand, the only allowed  nonrenormalizable term
in the superpotential with a leading contribution to the neutrino
mass matrix is of the type  $ W_{NR}=
L_iL_i\bar S_1\bar S_1/M_{NR}$\Ref\footZN{
A scenario that ensures large VEV's $<S>=<\bar S>=M_U$
is based on an interplay of soft
supersymmetry-breaking mass terms  of order $M_W$,
\ie , terms of the type $-M_W^2(|S_1|^2+|\bar S_1|^2)$
 and nonrenormalizable terms of the
type $W=\break
({\bf27}_I{\overline{\bf27}}_J)^K/M_{NR}^{2K-3}$, with $K\ge 4$.
This allows
for $M_U=
{\cal O}(M_W M_{NR}^{2K-3})^{1/{2K-2}}\geq {\cal
O}(M_U)\sim10^{16}$ GeV.
In order to prevent the terms in the $(S_1,\bar S_1) $
sector with $K<4$ one has to
impose a discrete symmetry; \eg , $Z_4$  symmetry
with $L_i$, $S_1$ and $\bar S_1$  transforming
with the  phases $ 2\pi/4$, $4\pi/4$ and $2\pi/4$,
respectively, ensures that the first nonzero term
in the $(S_1,\ \bar S_1)$ sector  corresponds to
$K=4$ while at the same time allowing  for the term $
L_iL_i\bar S_1\bar S_1/M_{NR}.$ Note, that this scenario
for the breakdown of the gauge symmetry at $M_U$ is again
motivated from the properties of string vacua.}.
This              modifies the neutrino mass matrix:
$$\left[\matrix{
              0&m\cr
              m&M_I\cr}\right],\eqn\nrm$$
where $M_I=M_U^2/M_{NR}$.
            The
%           $U(1)_{Y_1}\times U(1)_{Y_2}$
quantum numbers prevent  the contribution of any
non-renormalizable term to the
 $\nu \nu$ and $\nu L$
 masses  that would be of the
 order of  $M_U^K/M_{NR}^{K-1}$ for any $K>   1$.

As seen in the above model,
the gravity-induced
seesaw can be accommodated by
an interplay of the  renormalizable and
nonrenormalizable terms in
the superpotential, which takes place because
of a restricted representation of the Higgs sector.
While such a scenario is appealing on its own terms,
we would like to motivate its origin.
Study of the
effective Lagrangian of superstring vacua
provides a natural framework for  the
restricted representation of the chiral
supermultiplets.
One can also shed light on the
origin of the nonrenormalizable
terms in the superpotential, which in
string theory arise due to the exchange of
massive modes.

%There, the particle content of the matter supermultiplets
%is usually  constrained and thus at $M_U$
%the Yukawa couplings
%alone usually cannot provide for a large Majorana
%mass of the standard model singlet. However, the nonrenormalizable terms,
%which in string theory arise due to the exchange of massive modes,
%can in turn generate a large Majorana
%mass $M_I<M_U$
%for such a singlet.

 Let us first illustrate  the neutrino mass pattern
in the example discussed by Nandi and Sarkar\refmark\Nandi
with   a  gauge group $G\in E_6$  and
all the chiral supermultiplets (\ie ,
 those which contain
quarks, leptons and Higgs
particles) arising from $\bf 27$-plets of $E_6$.
The  renormalizable superpotential
is of the type
$
W_R={\bf27}_i{\bf27}_j{\bf27}_K$,
where quarks and leptons arise from ${\bf 27}_{i,j}$ and
Higgs  vacuum expectation values (VEV's) from
${\bf 27}_K$.
If one assumes  that all the exotic quarks acquire large masses
due to the  large
VEV's
%$M_{U1}$ and $M_{U2}$
 of the Standard Model  singlets $S_1$ and $S_2$,   while
the Higgs doublets  in ${\bf 27}_K$ account for the masses
of the ordinary
quarks and leptons,
this constrains\refmark\Nandi
the mass matrix of
the five neutral fermions $(\nu,N,N^c,\nu^c,L)$
to be  of the form:
$$
\left[\matrix{0&0&M_{U_1}&m_1&0\cr
              0&0&M_{U_2}&0&m_1\cr
             M_{U_1}&M_{U_2}&0&m_2&m_3\cr
             m_1&0&m_2&0&0\cr
             0&m_1&m_3&0&0\cr}\right].\eqn\renme$$
%See the table I  for
Under the
$SU(3)_L\times SU(3)_R\times SU(3)_C$
 subgroup of $E_6$, the
 neutral fields $ (\nu,N,N^c,\nu^c,L)$ are a part of a
${\bf (3,\bar 3, 1)}$  multiplet with the following entries:
$
\left[\matrix{N&\ \ &\ \ \cr
              \ \ &N^c&\nu\cr
             \ \ &\nu^c&L\cr}\right]$.
 $M_{U_1}$ and
 $M_{U_2}$ are masses due
to the  VEV's  of the   Standard Model singlets
%Higgs multiplets, which give mass to $-{1\over3}$ charged quarks
%($SU(2)_L$ singlet),
and $m_{1,2,3}$ are the light masses,
related to the
quark masses, which are due to the Standard Model Higgs
doublets.
The mass matrix \renme\       does not
have a large Majorana
mass along the diagonal of the lower two components.
There is a  heavy sector  with large masses $M_U$
%($ N^c$ and a linear combination of $\nu$ and $N$)  and
($ N^c$ and $(\nu + N)/\sqrt{2})$),
a light sector    with masses of order $m$
%        (mainly $ \nu^c,$
($(\nu - N)/\sqrt{2})$ and $(\nu^c - L)/\sqrt{2})$),
% and a linear combination of
%$\nu$ and $N$) as well as
and an ultra-light Standard Model singlet
%(mainly $L$)
($(\nu^c + L)/\sqrt{2})$)
with mass of order\refmark\Nandi\ $m^2/M_U$;
this pattern is
clearly  incompatible with experiment.

However,
nonrenormalizable terms could provide (along with
the renormalizable ones) a derived seesaw pattern.  Within
the $E_6$  gauge group there may be
 terms in the superpotential of the
type
$
W_{NR}=               {\bf27}_i{\bf27}_j{\overline{\bf27}_K}
{\overline{\bf27}_L}/M_{NR}
 $,
which can
account for the heavy Majorana masses\rlap.\refmark\Nandi
Namely, the mass matrix \renme\ becomes
$$
\left[\matrix{0&0&M_{U_1}&m_1&0\cr
              0&0&M_{U_2}&0&m_1\cr
             M_{U_1}&M_{U_2}&0&m_2&m_3\cr
            m_1&0&m_2&M_{I_1}&M_{I_2}\cr
            0&m_1&m_3&M_{I_3}&M_{I_4}\cr}\right]\eqn\nonrenmeII
$$
with $M_I=M_U^2/M_{NR}$.

%It can be shown\refmark\Nandi\
      Corrections to the
pattern \nonrenmeII\ are scaled by\refmark\Nandi\ $M_I/M_U\sim M_U/
M_{NR}$ and have been neglected\rlap.\Ref\footZZ{The physical light
doublet neutrino of mass $m\sim c m^2/M_I$ is
$\nu\cos\theta+N\sin\theta$ where $\tan\theta=M_{U_1}/M_{U_2}$,
and $\nu$ and $N$ are both doublets, while
the orthogonal combination has mass $\sim M_U$.  This does not cause any
problems with universality because there is exactly the same mixing
between the charged doublet partners.}
% Again the sufficient
%pattern to exhibit seesaw involves only three fields:
%\begin{equation}
%\left[\matrix{0&M_U&0\cr
%             M_U&0&m\cr
%             0&m&M_I\cr}\right]\label{eqfive}
%\end{equation}
%where $M_I$ is again reproduced from nonrenormalizable terms.
%The natural scale for $M_{NR}$ would be of order
%compactification scale,  $M_C$
%$M_{Pl}$ since
Such terms are expected to
 appear due to the exchange of heavy  modes.
Again $M_I$ is somewhat too large for, say, $M_U\sim2\times10^{16}$ GeV
and $M_{NR}\sim10^{18}$ GeV.
As we point out later,
in superstring theory
one can shed light on the
magnitude
of the nonrenormalizable terms of the superpotential
\REF\Dine{M. Cveti\v c, {\sl Phys. Rev. Lett.} {\bf59}, 1795 (1987).}
\REF\Lee{M. Cveti\v c, {\sl Phys. Rev.} {\bf D 37}, 2366 (1987);
M. Dine and C. Lee, {\sl Phys. Lett.} {\bf203B}, 371 (1988).}
in a quantitative way,\refmark{\Dine , \Lee}  and thus
account for an additional suppression factor.

The above example
is based on the constraints of the
 $E_6$ gauge group and the {\bf 27} representations.
In order
for       this scenario to be compatible
with the grand desert scenario of the  minimal
 supersymmetric  Standard Model,
 a number of  other
constraints
%of the supersymmetric string vacuum
have to be satisfied: {\it (i)} the gauge couplings have to
meet at $M_U\simeq (1-4)\times 10^{16}$ GeV, {\it (ii)} below
$M_U$ the gauge group has to be the Standard Model
group\Ref\dirpro{There could be additional group factors which commute
with the Standard Model.}, and
{\it (iii)} the particle content contributing to the
running of the gauge couplings has to be that of
the minimal supersymmetric Standard Model.

In superstring theories these constraints place conditions on the
string vacuum.
Perhaps the most difficult
to satisfy
(with no existing example available) is  {\it (iii).}
Generically, $(2,2)$ string vacua, {\it
e.g.,} Calabi-Yau manifolds with gauge and spin connection identified,
possess a large number of additional multiplets.
In particular, for vacua without Wilson lines, the gauge
group is
$E_6$, with {\bf27}'s, ${\overline{\bf27}}$'s, and {\bf1}'s of $E_6$.
%and  $\#{\bf 27}-\#{\overline{\bf27}}$ corresponding to
%the number of families.
Some of the particles in these multiplets
acquire large masses if there are flat directions in
the space of specific string vacua.  Finding flat
\REF\WittenII{E. Witten, {\sl Nucl. Phys. B} {\bf268,} 79 (1986).}
\REF\footII{This is one of the possible constructions of
$(0,2)$ superstring vacua.}
\REF\Cvetic{M.Cveti\v c, {\sl Phys. Rev. Lett.} {\bf59,} 2829 (1987).}
 directions\refmark{\WittenII ,\footII}
allows one to give large
VEV's  to fields in a
particular set of {\bf27}'s and ${\overline{\bf27}}$'s, which in
turn   can
give mass to some of the
unwanted massless multiplets. At the  same time,
 $E_6$ is broken down to
$SO(10)$ or $SU(5)$.
%The  natural  scale for $M_C$  is the compactification scale.
Explicit examples of such directions have been found in blown-up
\REF\Font{A. Font, L.~Iba\~nez, H.~P.~Nilles, and F.~Quevedo, {\sl Nucl.
Phys. B} {\bf307,} 105 (1988).}
\REF\Greene{ B. Greene, {\sl Phys. Rev.} {\bf D40}, 1645 (1989).}
\REF\Gepner{D. Gepner, {\sl Phys. Lett.} {\bf 199B}, 380 (1987).}
%\REF\Antoniadis{Antoniadis, low TeV..}
\REF\footIII{This is the case for all level 1 Ka\v c-Moody
algebra constructions of string vacua.}
orbifolds\refmark{\Cvetic ,\Font} as well as for a class of Calabi-Yau
manifolds\refmark\Greene\ based on Gepner's\refmark\Gepner\
 construction.

% for simply connected string vacua, the gauge group is a
%simple one : $E_6$ for (2,2) Calabi-Yau spaces
%and $SO(10)$ or  $SU(5)$ for (0,2) Calabi-Yau
%spaces which are  obtained by finding  corresponding
%flat directions.

Flat directions of (2,2) vacua
provide one with a new
class of ((0,2))
string  vacua. In such vacua
a large number of unwanted modes become heavy;
 however, the gauge group is still a simple GUT group
($SO(10)$ or $SU(5)$). Since  the
 matter supermultiplets are in the fundamental
representation or singlets of the gauge group\rlap,\refmark\footIII\
this prevents one from breaking the simple
GUT groups
down to the Standard Model, thus contradicting the constraint {\it
(ii).}

This  problem can
be remedied
by the  introduction of Wilson lines  on the compactified
space, allowing   a breakdown of the simple
gauge group ($E_6$, $SO(10)$, or $SU(5)$) to a direct product of
simple groups and $U(1)$'s.
It is in general
\REF\FontII{A. Font, L. Iba\~ nez, F. Quevedo and A. Sierra,
{\sl Nucl. Phys. }{\bf B331}, 421(1990).}
possible\refmark{\FontII,\Greene }
to introduce Wilson lines which  break the gauge group down to the
Standard Model. At the same time, this procedure decouples
a large number of unwanted modes.
Since there is no grand unification in the 4-dimensional theory
one does not expect observable proton decay, and
the relationship between  $V_{CKM}$ and $V_{\ell}$ is lost.

 Thus, a viable scenario %we could envision,
 which could satisfy constraints {\it (ii)} and {\it (iii)}
is to construct (2,2) string vacua with  flat directions
as well as  Wilson lines.
However, there  exists no explicit
construction of such a supersymmetric string vacuum which would contain
only the minimal Standard Model particle spectrum.

\REF\Kaplunovsky{V.~Kaplunovsky, {\sl Nucl. Phys.} {\bf B307}, 145
(1988)  and  erratum, to appear.}
The next issue to be addressed is the value of $M_C$,
which is the scale
at which the gauge couplings $g$, as determined at the tree
level of the string theory, are equal. $M_C$ is
determined\refmark{\Kaplunovsky} in the ${\overline {DR}}$ scheme by
the value of the Planck mass $M_{Pl}$
and of the gauge coupling $g$
in the following way:
\REF\FOOT{In literature, there are quoted
different values of $M_C$, differing by factors
of 2 or $\sqrt 2$.  As explained in erratum of Ref.~\Kaplunovsky ,
using the  correct  numerical form  of the original
formula\refmark\Kaplunovsky,  and the  tree level relation
 $g^2
=32\pi/(\alpha 'M_{Pl}^2)$, with $g$ defined according to the $GUT$
convention yields the  quoted result, which is the same as the
original numerical value in Ref.~\Kaplunovsky .
Note, that the gauge coupling
 $g$ as defined in the effective
string Lagrangian, \eg, P. Ginsparg, Phys. Lett.
{\bf B197}, 139 (1987), is
by a factor of $\sqrt 2$  smaller than the
gauge coupling $g$ defined according to the
GUT convention. Namely, in the effective string Lagrangian
the trace over the generators of the vector representation
of $SO(2N)$ gauge group
is chosen to be  $Tr(T^aT^b)=-2\delta^{ab}$,
while in the GUT theories
the convention is  $Tr(T^aT^b)=-\delta^{ab}$, thus rendring
the  gauge  coupling for a factor of $\sqrt2$
bigger in the latter case.}

$$
M_C={e^{(1-\gamma)/2}\sqrt2\over{3^{3/4}\sqrt{\pi\alpha '}
}}=g\times {e^{(1-\gamma)/2}\over{3^{3/4}4\pi}} M_{Pl}
=  g\times 0.043M_{Pl} =  g \times 5.2\times
10^{17}
\hbox{~GeV}.\eqn\mc$$
where $\gamma=0.57722$ is the Euler constant,
$g^2=32\pi/(\alpha 'M_{Pl}^2)$, with $g$ defined according to the $GUT$
convention\refmark\FOOT
and $M_{Pl}=1.2\times 10^{19}$GeV.

For the expected value  $g \sim 0.7$
this is one order of magnitude too large
compared with
$M_U\sim(1-4)\times
10^{16}$ GeV, which is the scale of the
gauge coupling unification  of the minimal supersymmetric Standard Model.
However, threshold effects, \ie, genus one corrections to the gauge
couplings, can split the gauge couplings at
$M_C$, thus in principle allowing for an effective unification scale
$M_U<M_C$.  Explicit calculations of the threshold
\REF\Louis{L.~Dixon, V.~Kaplunovsky, and J.~Louis, {\sl Nucl. Phys.}
{\bf B355},  649 (1991).}
\REF\Ross{L.~Iba\~nez, D.~L\"ust, and G.~Ross, {\sl Phys. Lett.}
{\bf B272}, 251 (1991); L. Iba\~nez and D. L\"ust, CERN
preprint, CERN-TH.6380/92.}
\REF\Dixon{L.~Dixon, unpublished.}
\REF\MSS{M. Cveti\v c, UPR--489--T, September 1991
 (unpublished).
}
%\REF\footIV{A detailed study should be done when Wilson lines
%are included.}
corrections for a class
of orbifolds are given in Refs. {\Kaplunovsky ,\Louis}.
Extensive study\refmark\Ross\ of  threshold corrections in
orbifolds indicate that
$M_U<M_C$ if the massless spectrum satisfies
certain constraints compatible with the
target space one-loop modular anomaly.
%There are few examples that satisfy this constraint.
In such examples one would obtain $M_U={\cal O}(e^{(-cR^2/\alpha ')})
M_C$ when the orbifold radius  is large
($R^2/\alpha '\gg 1$ ).
The positive coefficient $c$ depends\refmark\Ross\  on the modular
weights of the massless states.
As  we shall see later, the heavy Majorana
mass turns out to be
$M_I={\cal O}(e^{-c'R^2/\alpha ')})M_C$.
In order to ensure $M_U\sim 10^{16}$ GeV and
$M_I\sim 10^{12}$GeV , this in turn involves  detailed
constraints on coefficients   $c$ and $c'$.

In the following, we shall pursue a different approach, \ie ,
study of smooth Calabi-Yau spaces.
For simply connected (2,2)
Calabi-Yau manifolds  the nature of threshold
corrections is different.
Such spaces possess the
$E_6\times E_8$ gauge group, and the
compactified space corresponds to the  smooth
Calabi-Yau manifolds
with the radius of compactification being large, \ie ,
in the conformal field theory language this
corresponds to the (2,2) string vacuum with
large VEV's for {\it all} the moduli.
In this case the massive modes of the string theory do
not contribute to the threshold corrections\refmark\Dixon~,
and thus the gauge couplings do not have corrections
proportional to the powers of the moduli VEV's. Instead,
such corrections are milder, only logarithmic in
the VEV's of moduli. The threshold corrections for
 Calabi-Yau spaces  (with only one modulus\refmark\Dixon
as well as for arbitrary  number of \break
moduli\refmark\MSS) are of the
form:
$$
\Delta\left({16\pi^2\over
g^2}\right)\sim-b_G^{N=1}\ln(T+T^\ast),\eqn\thresh
$$
where the real part of the
$T$ field corresponds to
 an overall value of large moduli, {\it i.e.},
$T+T^\ast={\cal O}
(2R^2/\alpha')\gg 1$, where $R$ is the radius of the
compactification and $\alpha'$ is the string tension.
$b_G^{N=1}$ is  related to the one loop
 $N=1$ beta function $\beta_G$
  for $G=E_6$ or
$E_8$ as $\beta_G=b_Gg^3/16\pi$.
 Since threshold corrections \thresh\ to the gauge
coupling of each of the gauge groups
are proportional to the corresponding
$N=1$ beta function, this implies that
the slope of the running gauge couplings is not
changed. However the effective
gauge coupling  unification scale is
lowered:
$$
M_U^2={{M_C^2}\over{(T+T^*)}}=
{\cal O}({{M_C^2}\over{2R^2/\alpha'}})
\eqn\mu$$
The above results apply  only to simply connected
Calabi-Yau spaces. The gauge group   $E_6$
can be broken if Wilson lines are introduced. In this
case the contribution of the massless states to the
threshold corrections has not been studied, yet.
We proceed with the assumption that the
nature of the threshold corrections is still of the type \thresh .
 From \mu\ one then sees that for
$R^2/\alpha'\sim{\cal O}(20)$
the gauge unification scale is lowered
to $M_U \sim 6\times
10^{16}$ GeV, which is slightly too large. However,
equation \mu\ relates $M_U^2$  to  $R^2/\alpha '$
only by  orders of magnitude. Thus,
an additional factor
of $2$  in the relation of an overall modulus $Re
T$ to $R^2/\alpha '$
 enables  one to obtain $M_U$ in the preferred
range $4\times 10^{16}$ GeV.

We turn now to neutrino masses.
In particular, we would like to
address the size of the nonrenormalizable terms.
%  As seen from Eq.~\renme , $E_6$
%gauge symmetry in a $(0,2)$ theory obtained from a $(2,2)$ theory based
%on Yukawa couplings  \three\ cannot produce the desired large
%Majorana mass.  Note that due to the flat direction term in ${\bf27}_K$
%of Eq.~\ref{eqone} acquire VEV's of ${\cal O}(M_C)$.  However, with
%Wilson lines included, this pattern can still remain. The
%introduction of Wilson lines need not give large Majorana mass to
%$\nu^c$ and $L$.
%
%We turn now to nonrenormalizable terms of the type (\ref{eqfourp}).
 In
string theory the magnitude of the coefficient
%$1/M_{NR}$ is better
%understood; namely
 $M_{NR}$ is proportional to $M_C$.  However, one can
prove explicitly\refmark{\Lee}
that for all  $(0,2)$ string vacua
the nonrenormalizable terms are suppressed by an additional
factor $e^{-R^2/\alpha'}$, {\it i.e.,} the origin of the
nonrenormalizable terms is due only to worldsheet
instanton effects.
%Therefore, as the radius of the compactified space
%$R\rightarrow\infty$, the only terms surviving in the superpotential are
%renormalizable ones.
This is a
 general stringy result, proven
%\REF\CveticIII{M. Cveti\v c... blown-up orbifolds..}
explicitly on (blown-up) orbifolds\refmark{\Dine}
as well as
in sigma model
 perturbations\refmark{\Lee} of Calabi-Yau manifolds.
 Therefore:
$$
{1\over M_{NR}}={{{\cal O}(e^{-R^2/\alpha'})}\over M_C}.\eqn\mnr$$
By choosing vacuum expectation values along the flat direction
to be $M_C$ ( the only
 natural scale in the four-dimensional string vacuum)
 nonrenormalizable terms of the
type \nonrenmeII\  yield the heavy Majorana mass:
$$
M_I={M_C^2 \over M_{NR}}={\cal O}(e^{-R^2/\alpha'})M_C.\eqn\mi$$
It follows from
\mu\ that
we need $R^2/\alpha'={\cal O}(20)$ in order to achieve $M_U\sim10^{16}$
GeV. In this case
$M_I\sim10^{-8}M_C\sim10^{10}\hbox{~GeV}$.
Although these are only order of magnitude statements, it is
instructive to set the coefficients in \mu\ - \mi\
equal to unity. In that case the range $M_I \sim (4 \pm 3)
\times 10^{11}$ GeV suggested by the Solar neutrino deficit implies
$R^2/\alpha' \sim (13 - 15)$, yielding a slightly too large
$M_U \sim 7 \times 10^{16}$ GeV.

To summarize, the desired scale of the gauge coupling unification
 $M_U\sim 2 \times 10^{16}$ GeV and the scale of Majorana
neutrino masses
$M_I\sim4\times
10^{11}$ GeV, may be achieved \Ref\footMI{
There is a  less appealing  possibility in
   superstring theory, which for completeness we wish to discuss.
  In this case one would stick to
$(2,2)$ Calabi-Yau spaces  with Wilson lines and $R^2/\alpha'\sim{\cal
O}(1)$.  The threshold corrections are small and thus the
gauge coupling unification scale is $M_C\sim 4\times
10^{17}$GeV.
Because of the different matter representations, the
gauge couplings of the  semisimple group
({\it e.g.,} $SU(3)^3$)
factors  run
{\it differently} from $M_C$ to $M_U\sim10^{16}$ GeV,
where the semi-simple gauge group is broken down to the standard one.
Thus, the gauge couplings
are not equal at $M_U$ anymore.  In particular, for $SU(3)^3$ with $9q$
($6{\overline q}$) and $7\ell$
($4{\overline\ell}$) the difference in gauge
couplings is below current experimental uncertainties if $|M_C/M_U|<10$.
The known scenario for obtaining  large
$M_U$ is through an interplay of soft
supersymmetry-breaking mass terms (of order $M_W$)
 and nonrenormalizable terms of the
type $W=({\bf27}_I{\overline{\bf27}}_J)^K/M_{NR}^{2K-3}$, which
allow for
a large VEV of the Standard Model singlets  $\left<\phi\right>=
{\cal O}(M_W M_{NR}^{2K-3})^{1/{2K-2}}\geq{\cal
O}(M_U)\sim10^{16}$ GeV with $1/M_{NR}\sim{\cal O}(e^{-R^2/\alpha'})/M_C$
and $K\geq 4$.
 In addition, it is very difficult (as explored by
B. Greene, K. Kirklin, P. Miron and G. Ross,
{\sl Nucl. Phys. }{\bf B278}, 667(1986) and  {\bf B292}, 606 (1987))
 to decouple all
the unwanted particles at the same scale $M_U$.
Barring these problems, in this scenario one has
$1/M_{NR}={\cal O}(e^{-R^2/\alpha'})/M_C$ and $M_I=
(e^{-R^2/\alpha'})M_U^2/M_C$.
For $R^2/\alpha'\sim5$ and  $M_U\sim10^{16}$ GeV one obtains
 $M_I\sim10^{12}$ GeV.}
for a   superstring vacuum, corresponding to a
$(0,2)$
 Calabi-Yau obtained by  deforming a
 $(2,2)$  smooth large radius
 Calabi-Yau space along
the exactly flat directions.
The radius of  the
compactification  has to be in the range
$R^2/\alpha'={\cal O}(20)$, allowing
for $M_U^2=M_C^2/{\cal O}(2R^2/\alpha')$ and $M_I={\cal
O}(e^{-R^2/\alpha'})M_C$.
We believe it is intriguing
that one can obtain    $M_U$
and $M_I$ in the desired range on the basis of fairly
general stringy arguments.

We would like to thank R. Brustein, B. Greene and V. Kaplunovsky
for useful discussions and comments.
This research was  supported in part by the
U.S. DOE Grant DE-AC02-76-ERO-3071,
and by a junior faculty
SSC fellowship (M.C.).
\refout
\end